\documentclass[preprint,aps,12pt,showpacs,nofootinbib,tightenlines]{revtex4-1}
\usepackage{amsmath}
\usepackage{amssymb}
\usepackage{epsfig}
\usepackage{graphics}
\usepackage{color}

\begin{document}

\newcommand{\bea}{\begin{eqnarray}}
\newcommand{\eea}{\end{eqnarray}}
\def\beq{\begin{equation}}
\def\eeq{\end{equation}}

\def\pslash{\rlap{\hspace{0.02cm}/}{p}}
\def\eslash{\rlap{\hspace{0.02cm}/}{E}}
\let\jnfont=\rm
\def\NPB#1 {{\jnfont Nucl.\ Phys.\ B }{\bf #1} }
\def\PLB#1 {{\jnfont Phys.\ Lett.\ B }{\bf #1} }
\def\EPJC#1 {{\jnfont Eur.\ Phys.\ Jour.\ C }{\bf #1} }
\def\PRD#1 {{\jnfont Phys.\ Rev.\ D }{\bf #1} }
\def\PRL#1 {{\jnfont Phys.\ Rev.\ Lett.\ }{\bf #1} }
\def\MPLA#1 {{\jnfont Mod.\ Phys.\ Lett.\ A }{\bf #1} }
\def\JPG#1 {{\jnfont J.\ Phys.\ G }{\bf #1} }
\def\CTP#1 {{\jnfont Commun.\ Theor.\ Phys.\ }{\bf #1} }
\def\JHEP#1 {{\jnfont JHEP \ }{\bf #1} }
\def\NPPS#1 {{\jnfont Nucl.\ Phys.\ Proc.\ Suppl.\ }{\bf #1} }
\def\btt#1{{tt$\backslash$#1}}


\title {Scalar Septuplet Dark Matter and Enhanced $h\rightarrow \gamma\gamma$  Decay Rate}
\author{Yi Cai $^a$$^{,}$$^{e}$}
\email{yicai@sjtu.edu.cn}
\author{Wei Chao $^{a}$$^{,}$$^{b}$$^{,}$$^{e}$}
\email{chaow@physics.wisc.edu}
\author{Shuo Yang $^{c}$$^{,}$$^{d}$$^{,}$$^{e}$}
\email{yangshuo@lnnu.edu.cn}
\affiliation{$^a$ INPAC, Shanghai Jiao Tong University, Shanghai, P.R. China
\\
$^{b}$ Department of Physics, University of Wisconsin-Madison, Madison, WI 53706, USA\\
 $^{c}$ Physics Department, Dalian University, Dalian, 116622, P.R. China\\
$^{d}$ Center for High Energy Physics, Peking University, Beijing, 100871, P.R. China\\
$^{e}$ Kavli Institute for Theoretical Physics China, CAS, Beijing 100190, P.R. China}

\begin{abstract}
Inspired by recent results on the Higgs search from ATLAS and CMS, we extend the SM with complex septuplet scalars. The lightest neutral component of the septuplets is a natural cold dark Matter candidate and the charged components can contribute to the $h\to \gamma \gamma $ decay rate, providing a significant enhancement factor. The dark matter phenomenology and possible collider signatures of the model are investigated. We find a dark matter candidate with mass around 70 GeV consistent with astrophysical and direct detection constraints. In the meanwhile, the enhancement factor of $h\to \gamma \gamma $ decay rate can be in the range $1.5\sim 2$.

\end{abstract}
\pacs{ Higgs, Dark Matter, Collider Phenomenology} \maketitle

\section{ Introduction}

In the Standard Model (SM), the Higgs mechanism provides an explanation to the spontaneous electroweak symmetry breaking, but
the Higgs boson itself left no trace in all the previous high-energy collider experiments. Recently both CMS and ATLAS collaborations have announced an observation of a Higgs-like boson at about $5\sigma$ confidence level, which is a milestone for fundamental physics.

It is quite interesting to note that both CMS and ATLAS collaborations show a significant enhancement in the diphoton channel~\cite{ATLAS,CMS}
\begin{eqnarray}
{\left[\sigma(gg\to h) \times {\rm BR} (h\to \gamma\gamma) \right]_{\rm obs}\over \left[\sigma(gg\to h)\times {\rm BR} (h\to \gamma \gamma)\right]_{\rm SM }} \in [1.5,~ 2.0 ],
\end{eqnarray}
while $\sigma (gg\to h) \times {\rm BR} (h \to ZZ^*) $ and $\sigma(gg\to h) \times (h \to WW^*) $ seem compatible with the SM prediction.
In the absence of direct signals of new physics at colliders, the enhancement of $h\rightarrow \gamma\gamma$ rate is an
especially important hint of the underlying new physics, which is possibly a result of loop contributions of
new physics. Extensive studies of the enhancement in the extended scalar frame~\cite{hrr,hrrHTM,hrrHTM2,hrrHTMWang}
have been carried out.
On the other hand, some other observations also point to physics beyond the SM. Precise cosmological observations have confirmed the existence of non-baryonic cold dark matter. It would be quite appealing if we can solve two problems together in one certain minimal extension of the SM.

In conventional dark matter models, a particle can be long lived stable dark matter candidate due to certain discrete symmetry, such as $Z_2$ symmetry, $R$-parity in the minimal supersymmetric standard model (MSSM)~\cite{SUSYMDM} and $K$-parity in the universal extra dimension models \cite{UEDDM}. It can also be stable due its special representation under the SM gauge group, such as minimal dark matter model \cite{MDM}. In this work, we extend the SM with two colorless complex scalar septuplets providing a natural dark matter candidate and accommodating the enhancement factor in the decay rate of $h\to \gamma\gamma$. The dark matter phenomenology and possible collider signatures of the model are investigated.


The paper is organized as followings: In section II we describe the model in detail. Section III is devoted to the study of the dark matter phenomenology. In section IV, We study the contribution of new charged particles to $h\to\gamma\gamma$ decay. The collider phenomenology is discussed in section V. The last part is the concluding remarks.





\section{The model}
We extend the Standard Model with two colorless complex scalar septuplets, which can be written in
component as
\begin{eqnarray}
H_7&=&\left(
\begin{array}{lllllll}
H_p^{+++}&H_p^{++}&H_p^{+}&{(H^{0}+i\ A^{0})/ \sqrt{2}}~&H_m^{-}~&H_m^{--}&H_m^{---}
\end{array}
\right)^T\\
\Phi_7&=&\left(
\begin{array}{lllllll}
\Phi_p^{+++}& \Phi_p^{++}& \Phi_p^{+}~~& {(\Phi_R^{0}+i\ \Phi_I^{0} )/ \sqrt{2}}~~& \Phi_m^{-}~& \Phi_m^{--}& \Phi_m^{---}
\end{array}
\right)^T.
\end{eqnarray}
The Lagrangian of the scalar sector is
\beq
\label{eq:lagrangianmultiplet}
{\cal L} = \left(D_\mu H_7 \right)^\dagger \left(D^\mu H_7\right)+\left(D_\mu \Phi_7 \right)^\dagger \left(D^\mu \Phi_7\right)-V(H_7, \Phi_7,\Phi) \quad ,
\eeq
where $\Phi$ is the SM Higgs doublet. The covariant derivative is given by
\beq
D_\mu \equiv \partial_\mu -i g \tau^{(n)}_a W_\mu^a-i g_Y Y B_\mu \quad ,
\eeq
where $\tau_a^{(n)}$ stands for the $n$ dimensional representation of the
$SU(2)_L$ generators and $Y$ is the hypercharge.
The most general renormalizable scalar potential is given by
\begin{equation}
\label{eq:potentialmultiplet}
\begin{split}
V(H_7,\Phi_7, H) &= \mu^2\Phi^{\dagger} \Phi+\lambda_1(\Phi^{\dagger} \Phi)^2+ \mu_1^2 H_7^\dagger H_7 +
\lambda_2 \left(H_7^\dagger H_7\right)^2 +\mu_2^2 \Phi_7^\dagger \Phi_7^{} +\lambda_3^{} \left(\Phi_7^\dagger \Phi_7^{}\right)^2\\
&+ \lambda_4
\left(\Phi^{\dagger} \Phi\right) \left(H_7^\dagger
H_7\right)+ \lambda_5 \left(H_7^\dagger \tau_a^{(7)} H_7\right)^2
+ \lambda_6 \left(\Phi^\dagger \tau_a^{(2)} \Phi\right)
\left(H_7^\dagger \tau_a^{(7)} H_7\right)\\
&+ \lambda_7
\left(\Phi^{\dagger} \Phi\right) \left(\Phi_7^\dagger
\Phi_7\right)+ \lambda_8 \left(\Phi_7^\dagger \tau_a^{(7)} \Phi_7\right)^2
+ \lambda_9 \left(\Phi^\dagger \tau_a^{(2)} \Phi\right)
\left(\Phi_7^\dagger \tau_a^{(7)} \Phi_7\right)\\
&+\left\{  \mu_3^2 (\Phi_7^{} \hat C H_7^{}) +\lambda_{10}^{} (\Phi_7^{} \hat C H_7^{}) (\Phi^{\dagger} \Phi)+\lambda_{11}^{} (\Phi_7^{} \hat C H_7^{}) (H_7^\dagger H_7) \right. \\
&+\left. \lambda_{12} (\Phi_7^{} \hat C H_7^{}) (\Phi_7^\dagger \Phi_7^{}) +\lambda_{13} (\Phi_7^{} \hat C H_7^{})^\dagger (\Phi_7^{} \hat C H_7^{})+\lambda_{14} (\Phi_7^{} \hat C H_7^{})^2 +h.c. \right\}\; ,
\end{split}
\end{equation}
where $a$ sums over $1,2,3$. $\lambda_{10} \sim \lambda_{14}$ and $\mu_3^2$ are complex parameters. $\hat C$ is a $7\times7$ matrix
\beq
\hat C=\left(
\begin{array}{ccccccc}
0&0&0&0&0&0&-1\\ 0&0&0&0&0&1&0\\ 0&0&0&0&-1&0&0 \\ 0&0&0&1&0&0&0 \\ 0&0&-1 &0&0&0&0\\0&1&0&0&0&0&0\\-1&0&0&0&0&0&0 \end{array}
\right).
\eeq
We name this model S7M, short for the Standard Model with scalar septuplets.
According to the scalar potential given in Eq.~(\ref{eq:potentialmultiplet}),
we can calculate the mass eigenvalues of all the scalar components.
Here we assume $\mu_1^2 \ll \mu_2^2 $ for simplification, such that the mass matrix of septuplets can be
diagonalized easily without loss of generality.
We need two septuplets to generate a large mass splitting between the charged and neutral components of $H_7$.
In the low energy phenomenological analysis $\Phi_7$ actually decouples.
If we ignore the last two lines in Eq.~(\ref{eq:potentialmultiplet}),
the mass eigenvalues of two septuplets can be written as
\begin{eqnarray}
M_{H_7^Q}^2 &=& \mu_1^2 + {1 \over 2} \lambda_4^{} v^2 +{1\over 2 }\lambda_6  (Q/e) v^2 \; , \\
M_{\Phi_7^Q}^2& =&  \mu_2^2 + {1 \over 2} \lambda_7^{} v^2 +{1\over 2 }\lambda_9  (Q/e) v^2 \; ,
\end{eqnarray}
where $Q$ is the electric charge of the component field.
Obviously the neutral component of $H_7$ is not the lightest one in this case and can not be the dark matter candidate.
Here we can set $\lambda_6^{}, \lambda_9^{} \sim 10^{-6} $, such that the contributions of these terms can be neglected compared with loop-level corrections to the scalar masses. Thus the mass splitting between the charged and neutral components can be of several hundred ${\rm MeV}$ \cite{SDM}.

With the contribution from the terms in the fourth line of Eq.~(\ref{eq:potentialmultiplet}), the mass eigenvalues of $H_7$ can be approximately written as
\begin{eqnarray}
M_{H_{p,m}^Q}^2 &\approx& \mu_1^2 + {1 \over 2} \lambda_4^{} v^2  -{\left(Re[\mu_3^2] ^2+ Im[\mu_3^2]^2 \right)  \mu_2^{-2} } \; , \\
M_{H^0}^2  &\approx& M_{H_{p,m}^Q}^2+ 2 Re[\mu_3^2] Im[\mu_3^2] \mu_2^{-2} \; , \\
M_{A^0}^2  &\approx& M_{H_{p,m}^Q}^2- 2 Re[\mu_3^2] Im[\mu_3^2] \mu_2^{-2} \; . \label{eq10}
\end{eqnarray}
Here we ignored the contribution of  the term proportional to $\lambda_{10} $ for simplification, since the contribution of this term can be absorbed into the term proportional to $\mu_3^2$.
So either the CP-even or the CP-odd neutral component of the $H_7$ is the lightest one.
Without loss of generality, we assume the CP-even component is the lightest and play the role of dark matter.
In the following sections, we study the dark matter phenomenology of $H^0$ and the implication of the
charged components of $H_7$ in the enhancement of $h\rightarrow \gamma\gamma$ decay rate.
There is an extensive study on scalar multiplet dark matter including the septuplet case in Ref.\cite{SDM}.
In this work, with the presence of the scalar potential, a new viable region of low mass dark matter is open
and we extend the studies of scalar septuplet model to Higgs decay and collider phenomenology of new scalars at the LHC.

We also study  the constraints implied by septuplet loop contributions to electroweak precision
observables (EWPO), which can be characterized by the leading effects of these corrections
in terms of the oblique parameters \cite{ewprecision}.  Since we have assumed $\Phi_7$ is heavy
and nearly degenerate, so we only need to consider the contributions of $H_7$ to the gauge boson
self-energy functions,
which can be expressed as
\begin{eqnarray}
\alpha T& \approx& \sum_{Q=-2}^{3}{ e^2 \over \hat s^2 M_W^2} F_{_Q, _{Q-1}} {1 \over 24 \pi^2 } {(\Delta m_{ _Q, _{Q-1}}^2 )^2 \over M_{H_{p,m}^Q}^2 + M_{H_{p,m}^{Q-1}}^2} \; , \\
\alpha S &=& 0 \; ,
\end{eqnarray}
where $\hat s$ is the sine function of the weak mixing angle in the $\overline{MS}$ scheme,
$\Delta m_{ _Q, _{Q-1}}^2 =M_{H_{p,m}^Q}^2 - M_{H_{p,m}^{Q-1}}^2$ and $F_{_Q, _{Q-1}}$ is a factor
induced by $SU(2)_L$ generators.
To be specific, we have $F_{_3, _2} = F_{_{-2},  _{-3}}=6$, $F_{_1, _1} = F_{_{-1},  _{-2}}=10$
and $F_{_1, _0} = F_{_{0},  _{-1}}=12$.
As stated above, we only consider the mass splitting between the neutral components
and charged components.   The latest global fit to the EWPO yields  $S=-0.1 \pm 0.10 (-0.08)$
and $T=-0.08\pm 0.11 (+0.09)$ \cite{pdg}.
The most important constraint from the T parameter gives us a $1\sigma $  lower bound
of the mass splitting bewteen the neutral and charged components of $H_7$,
about $42~{\rm GeV}$, by assuming $M_{H_{p,m}^Q} \sim 100 ~{\rm GeV}$.


\section{Dark Matter}

\begin{figure}[htb]
\begin{center}
 \includegraphics[width=5in]{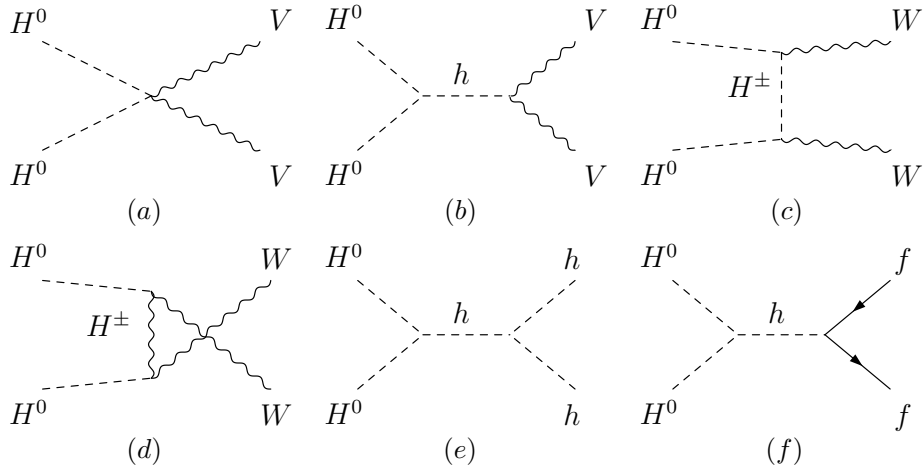}
\end{center}
\caption{Feynman diagrams contributing to dark matter annihilation. $V$ represents $W$ or $Z$ boson.}
\label{Fig:Dm}
\end{figure}
The fact that about 23\% of the Universe is made of dark matter
has been firmly established, while the nature of the origin of dark matter still elude us.
A weakly Interacting massive particle (WIMP) is a promising dark matter candidate,
since the WIMP relic density can be naturally around the experimental value \cite{wmapp}
\begin{eqnarray}
\Omega h^2 =0.1123\pm 0.0035
\end{eqnarray}
for a WIMP mass around the electroweak scale.
As was shown in section II, $H^0$ is the lightest stable particle in our model
and thus can be the cold dark matter candidate.
It couples to the gauge bosons as in the minimal dark matter model
and to the SM Higgs boson as well in this model.
We show in Fig. 1 the Feynman Diagrams for the dark matter annihilation.
In the high dark matter mass region, where the dark matter mass is bigger than the SM Higgs boson mass $M_h$,
processes shown in Fig.~\ref{Fig:Dm} $(a)-(e)$ dominate the annihilation of the dark matter.
For  $M_W< M_{H^0} < M_h$,  processes shown in Fig.~\ref{Fig:Dm} $(a)-(d)$ dominate the annihilation of $H^0$.
In this case,  there may be cancellations between  Fig.~\ref{Fig:Dm} $(a)-(c) $ and Fig.~\ref{Fig:Dm} (d)
in certain parameter space \cite{InertDM}.
For ${1\over2}  M_h < M_{H^0} < M_W$, Fig.~\ref{Fig:Dm} $(f)$ dominates the annihilation of the dark matter.
The dark matter mass is chose to be larger than one half of the Higgs mass to avoid Higg invisible decay.

\begin{figure}[htb]
\begin{center}
 \includegraphics[scale=0.55]{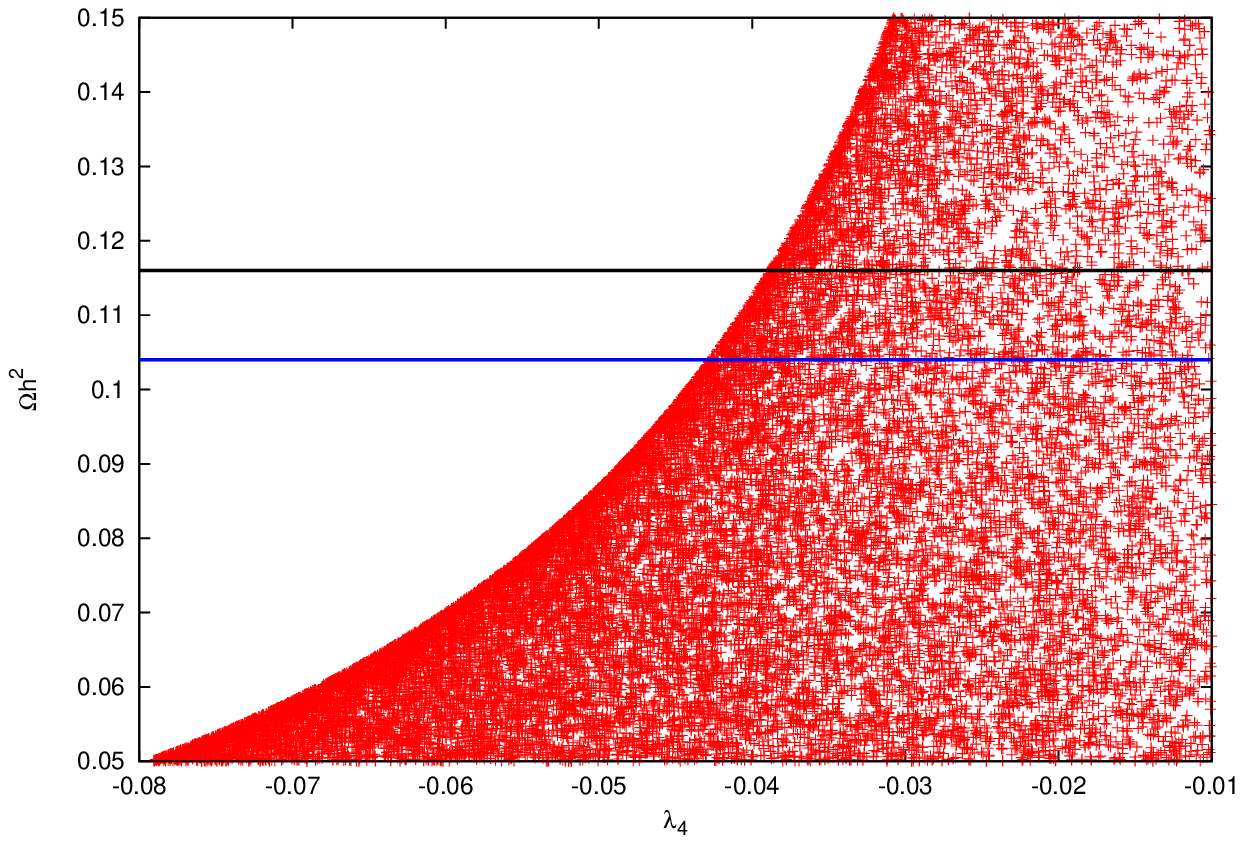}
  \includegraphics[scale=0.55]{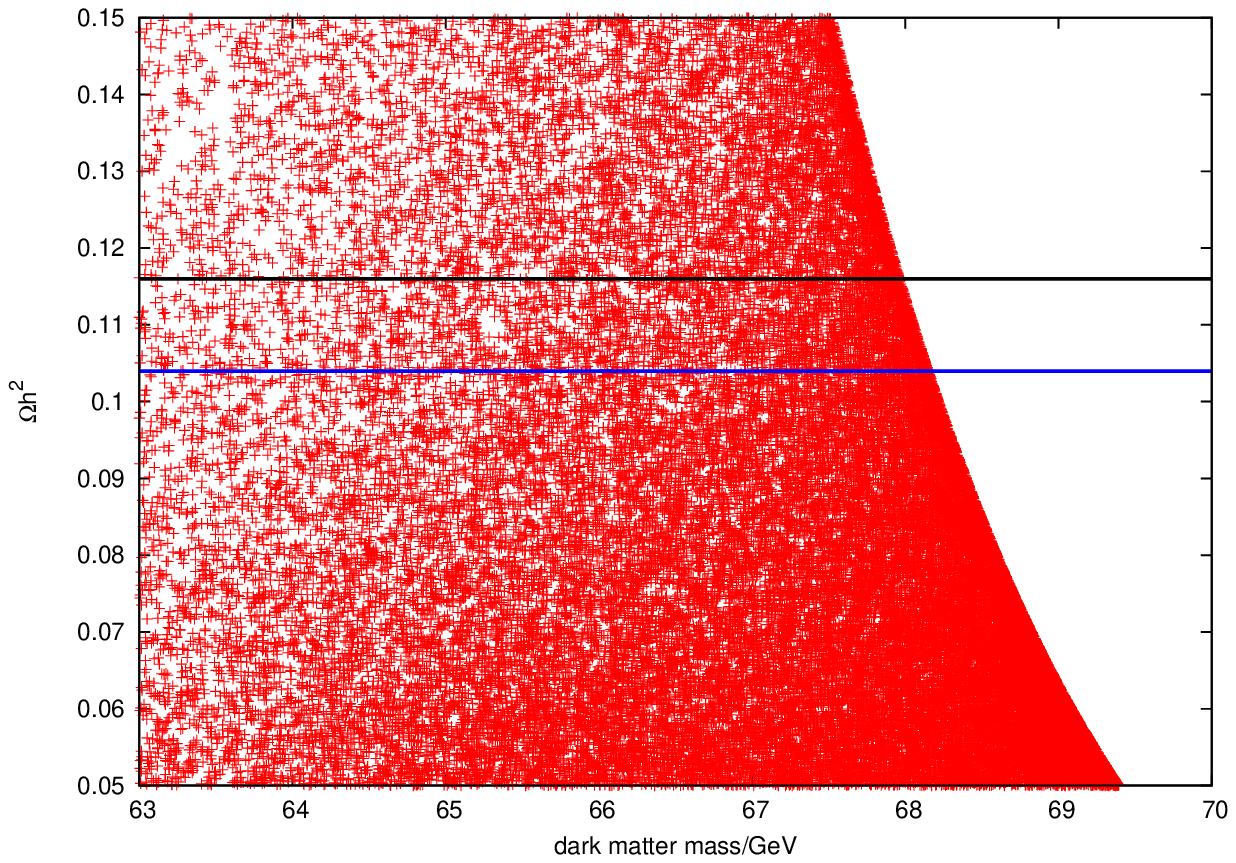}
\end{center}
\caption{ Left panel: The dark matter relic abundance versus $\lambda_4$ by varying dark matter mass in the range$[ {1\over 2} M_h^{},~ M_W^{}]$,
with the horizontal band being the observed value of dark matter relic density;
Right panel: The dark matter relic abundance versus $M_D$ by Varying $\lambda_4$ in the range $[-0.1, -0.01]$.}
\label{dmrelic}
\end{figure}

The relic density and the direct detection cross section are calculated with MicrOMEGAs~\cite{MicO},
which solves the Boltzman equation numerically and utilizes CALCHEP~\cite{calchep}
to calculate the relevant cross sections.
In the left panel of Fig.~\ref{dmrelic}, we plot the dark matter relic abundance
as a function of $\lambda_4$ by varying dark matter mass  $M_D$ in the low mass region
$M_D \in [ {1\over 2} M_h^{}, M_W^{}]$. The horizontal band represents the $3\sigma$ region consistent
with current relic density measurement from WMAP, $0.104< \Omega h^2 <0.116$. The SM Higgs mass is chosen to be
$m_h=125$ GeV. In this case the $H^0 H^0 \to \bar b b$ via the exchange of a SM Higgs is the dominant annihilation channel of the dark matter. The annihilation cross section is roughly proportional to $\lambda_4^2$.
It is easy to see from Fig.~\ref{dmrelic} that $\lambda_4$ should be be larger than about $-0.04$
to get the correct dark matter relic abundance.
In the right panel of Fig. \ref{dmrelic}, we plot the dark matter relic density as
a function of $M_D$ by varying $\lambda_4$ in the range $[ -0.1,~-0.01]$, where the enhancement of
$h\rightarrow \gamma\gamma$ rate is sizable,which will be elaborated in the next section.

\begin{figure}[htb]
\begin{center}
 \includegraphics[scale=0.8]{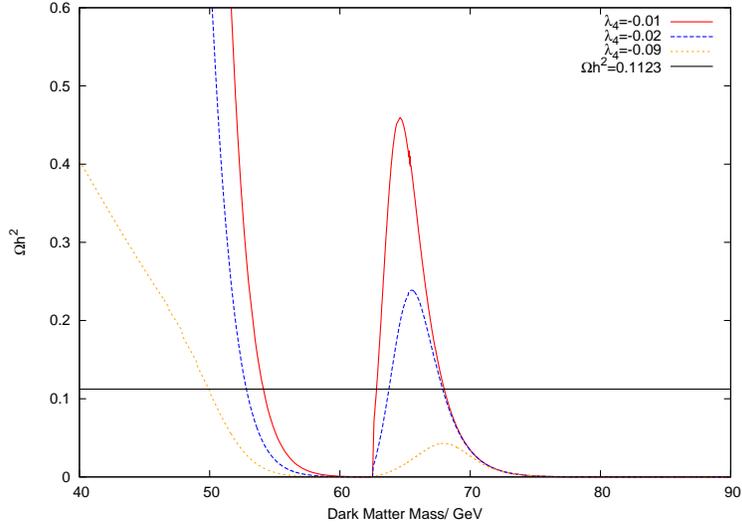}
\end{center}
\caption{ The dark matter relic density as a function of dark matter mass for typical values of $\lambda_4$. The horizontal line represents the central value of the observed relic density. }
\label{DMfixlam}
\end{figure}

In Fig.~\ref{DMfixlam}, we plot the dark matter relic density as a function of dark matter mass for $\lambda_4=-0.01,\ -0.02,\ -0.09$. In the small mass region $m_{H^0}<60$ GeV, the cross section is dominated by the process $H^0H^0 \to h^* \to f\bar{f}$. As the dark matter mass $m_{H^0}$ approaches $m_h/2$, the annihilation cross section is resonantly enhanced, so that the relic density is significantly suppressed. Similarly, when $m_{H^0}$ approaches $m_W$, the annihilation is dominated by $H^0H^0 \to h \to WW^*$.

\begin{figure}[htb]
\begin{center}
 \includegraphics[scale=0.6]{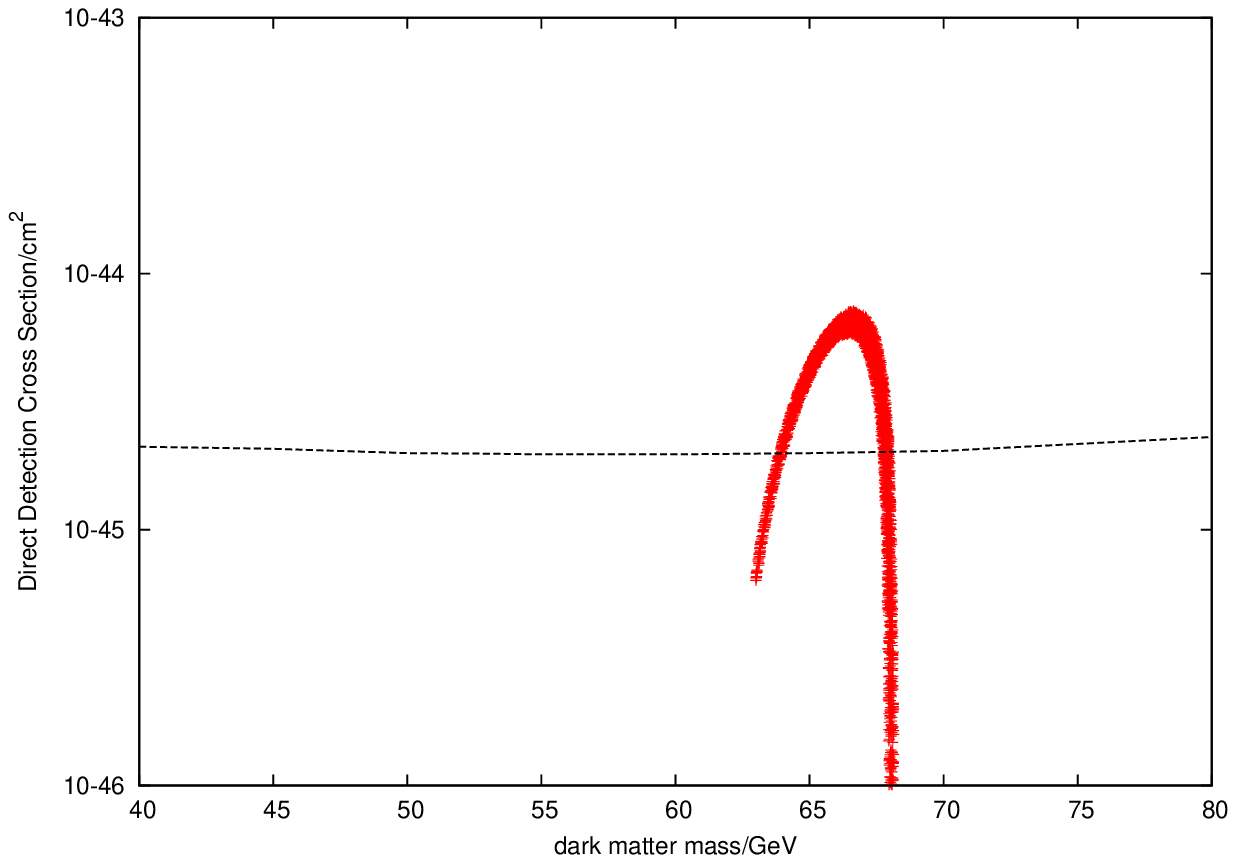}
 \includegraphics[scale=0.6]{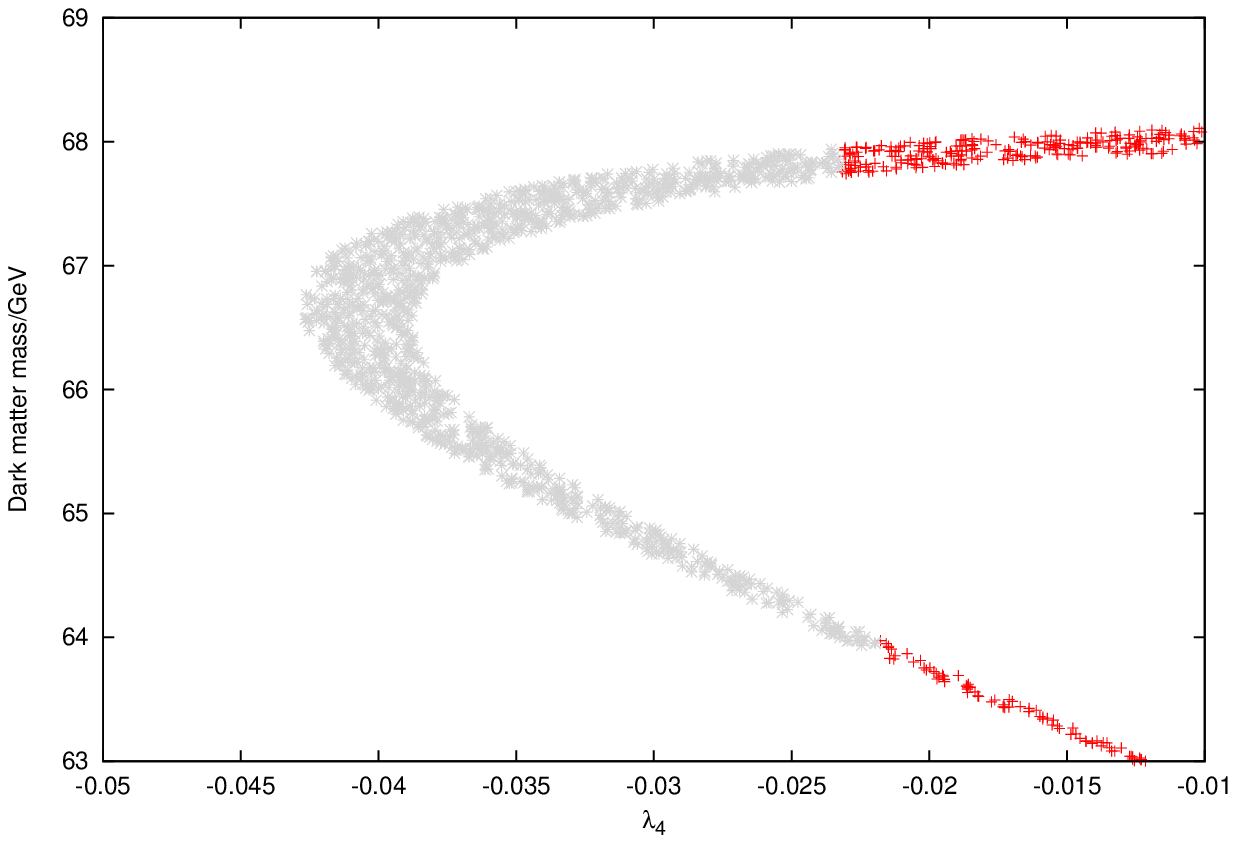}
\end{center}
\caption{ Left panel: Dark matter nucleus scattering cross section versus dark matter mass. The grey line is the current bound from Xenon100;
Right panel: Dark matter mass versus $\lambda_4$ constrained by the dark matter relic abundance and dark matter direct detection.}
\label{dmdirect}
\end{figure}

In the left panel of the  Fig.~\ref{dmdirect},
we plot the cross section of dark matter scattering off nucleus
as a function of dark matter mass by varying $\lambda_4$ in the range $[-0.1,~-0.01]$,
where all the red points are consistent with the dark matter relic density.
The grey line is the current bound of XENON 100 released in 2012~\cite{XENON}.
In the right panel of Fig.~\ref{dmdirect}, we show the viable parameter space where the model gives
the right dark matter relic abundance and the direct detection cross section is below the XENON 100 exclusion limit.
To be consistent with the latest XENON 100 result in the $2\sigma$ level, $\lambda_4>-0.043$ and
$63\ {\rm GeV}<M_{H^0}<68\ {\rm GeV}$. The grey points are above the central value of the limit while the red ones
are below the limit.

\section{ The Higgs Decay $h\rightarrow \gamma \gamma$ }
 \noindent
ATLAS and CMS collaborations have released their results on the Higgs search based on 2012
data \cite{ATLAS,CMS}, 6 fb$^{-1}$ at 8 TeV, and the 2011 data, 5 fb$^{-1}$ at 7 TeV \cite{ATLAS2011,CMS2011}.
Both collaborations have observed a Higg boson like particle.
Another intriguing result is the enhancement of $h\rightarrow \gamma\gamma$ decay rate,
about $1.5 \sim 2$ larger than the SM prediction \cite{ATLAS,CMS,ATLASrr,CMSVBF}.
Based on recent search results, studies of global fit~\cite{Hfit} show
that some new theoretical proposal, which predict more $\gamma \gamma$ events,  are favored over the SM.
Inspired by the enhancement of $h\rightarrow \gamma\gamma$, extensive studies have been carried out
recently in extended scalar frameworks~\cite{hrr,hrrHTM,hrrHTM2,hrrHTMWang}.

In the S7M, there are additional contributions to the decay width $\Gamma(h\to
\gamma\gamma)$
from charged scalars $H_{p,m}^{\pm}$, $H_{p,m}^{\pm\pm}$, $H_{p,m}^{\pm\pm\pm}$
\footnote{The loop induced $h\rightarrow Z\gamma$ would also receive contributions from charged scalars.
However, we will not consider this channel because it is not expected to be observable in the early LHC stage.}
in the loop, which could accommodate the enhancement of $h\rightarrow \gamma\gamma$ as presented below.
The decay width $\Gamma(h\to
\gamma\gamma)$ in the S7M can be expressed as \cite{Hreview2}
\begin{eqnarray}
\nonumber
\Gamma(h\to \gamma\gamma) &=& \frac{\alpha^2
  m^3_{h}}{256\pi^3v^2}\Bigg|  F_{1}(\tau_W)+ \sum_{i} N_{cf} Q^2_{f} y_{f} F_{1/2}(\tau_f)\\
&+&  g_{_{H_{p,m}^{\pm}}}F_{0}(\tau_{H_{p,m}^\pm})+  4g_{_{H_{p,m}^{\pm\pm}}}F_{0}(\tau_{H_{p,m}^{\pm\pm}})+9g_{_{H_{p,m}^{\pm\pm\pm}}}F_{0}(\tau_{H_{p,m}^{\pm\pm\pm}})
\Bigg|^2 ,
\label{gamrr}
\end{eqnarray}
where $\tau_i=\frac{4m_i^2}{m_h^2}$ and the couplings are
\bea \nonumber
g_{_{H_{p,m}^{\pm}}}=-\frac{v}{2m_{H^\pm}^2}g_{hH^+ H^-},
~~g_{_{H_{p,m}^{\pm\pm}}}=-\frac{v}{2m_{H^{\pm\pm}}^2}g_{hH^{++}
H^{--}},~~g_{_{H_{p,m}^{\pm\pm\pm}}}=-\frac{v}{2m_{H_{p,m}^{\pm\pm\pm}}^2}g_{hH^{+++}
H^{---}}\eea with
\beq
g_{hH^{+}H^{-}}=g_{hH^{++}H^{--}}=g_{hH^{+++}H^{---}}=-\lambda_4v .
\eeq
In the simplest case, the masses of the charged scalars are nearly degenerate.
Here $N_{cf}$ and $Q_f$ are the color factor and the electric
charge of the fermion $f$ running in the loop respectively.
The dimensionless loop factors \cite{HHG} are
\begin{eqnarray}
F_1 = 2+3\tau + 3\tau(2-\tau)f(\tau), \quad F_{1/2} =
-2\tau[1+(1-\tau)f(\tau)], \quad F_0 = \tau[1-\tau f(\tau)],
\label{ffun}\end{eqnarray}
with
\begin{equation}
f(\tau) = \left\{ \begin{array}{lr}
[\sin^{-1}(1/\sqrt{\tau})]^2, & \tau \geq 1 \\
-\frac{1}{4} [\ln(\eta_+/\eta_-) - i \pi]^2, & \, \tau < 1
\end{array}  \right.\ ,
\end{equation}
where $\eta_{\pm} = 1 \pm \sqrt{1-\tau}$.

We define a simple ratio
\begin{equation}
\mu_{\gamma\gamma}=\frac{(\sigma(gg \rightarrow h)
\times {\rm BR}(h\rightarrow \gamma\gamma))^{S7M}}
{(\sigma(gg \rightarrow h)\times
{\rm BR}(h \rightarrow \gamma\gamma))^{SM}} \, .
\label{eq:ratio}
\end{equation}
In the S7M, we don't introduce any sources
that change the Higgs production rate through gluon fusion.
Thus the ratio $\mu_{\gamma\gamma}$ can be further simplified as $\mu_{\gamma\gamma}=\frac{h\rightarrow \gamma\gamma^{S7M} }{h\rightarrow \gamma\gamma^{SM}}$.

\begin{figure}[htb]
\begin{center}
 \includegraphics[scale=0.6]{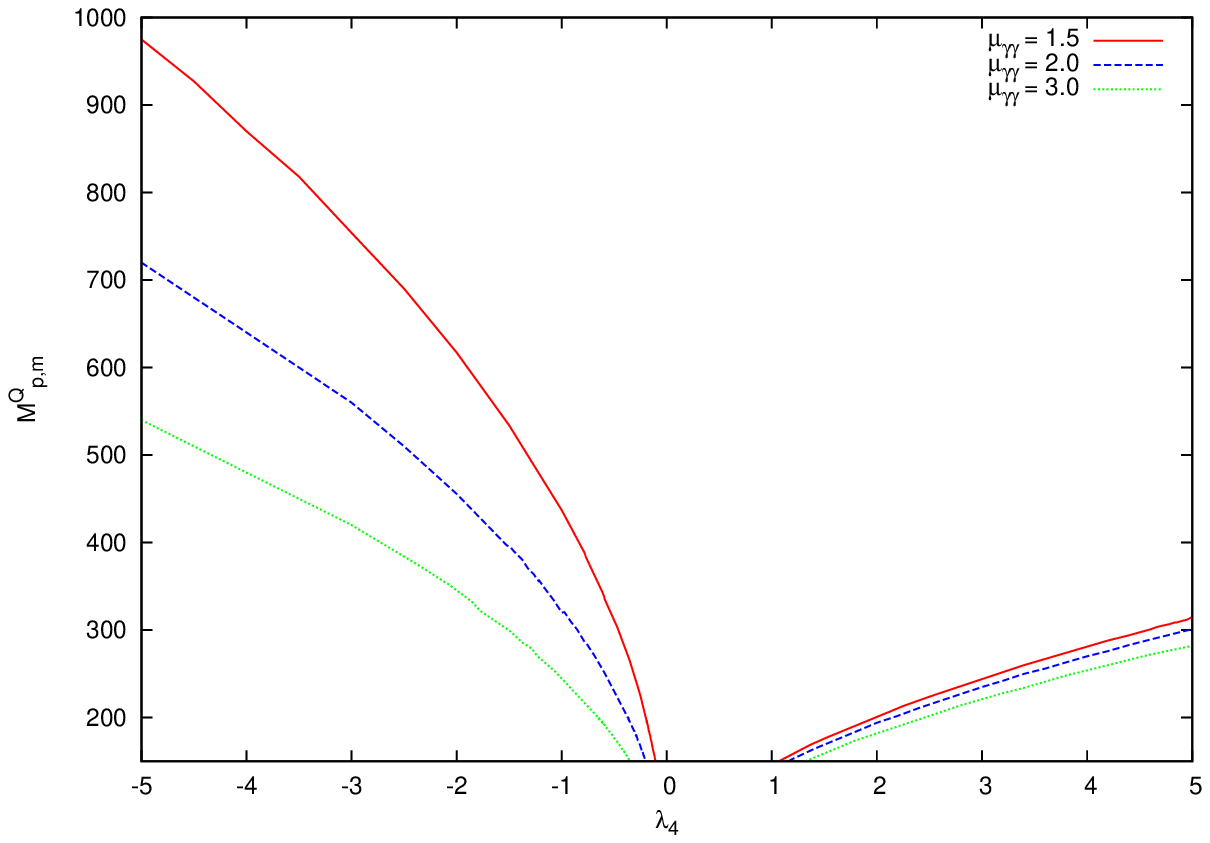}
  \includegraphics[scale=0.6]{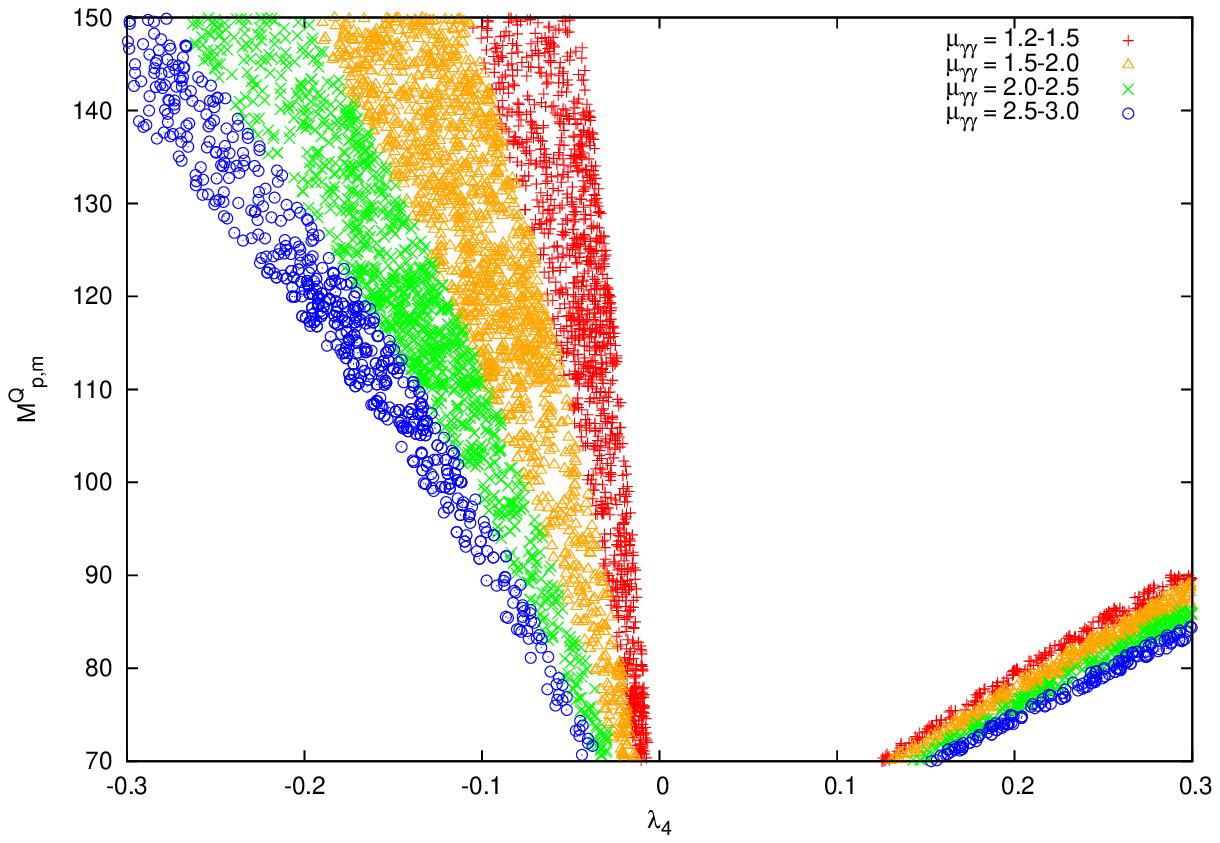}
\end{center}
\caption{The ratio $\mu_{\gamma\gamma}$ in the plane of [$\lambda_4$,$m_{H_{p,m}^{Q}}$] for $-5<\lambda_4<5$ (left panel) and $-0.3<\lambda_4<0.3$ (right panel). In the figures, all the components are assumed to be degenerate and the Higgs mass is fixed to be 125 GeV.}
\label{Fig:hrr}
\end{figure}

The Higgs decay rates of the di-photon and other channels are calculated
with modified public package HDECAY\cite{Hdecay}.
In this model, Higgs does not decay invisibly to a dark matter pair, which is not favored by the data~\cite{Hfit},
since opening the invisible channel will reduce the visible rates
and lead to a tension with the observed rates in the di-photon and $ZZ^*$ channels.

In Fig.~\ref{Fig:hrr}, we show the ratio $\mu_{\gamma \gamma}$ in the plane of [$\lambda_4$, $m_{H_{p,m}^{Q}}$]. The
$\lambda_4$ and $m_{H_{p,m}^{Q}}$ are treated as free parameters in the range:
\beq
-5< \lambda_4<5;\quad 70\ \rm{GeV} <m_{H_{p,m}^Q}<1000\ \rm{GeV}
\eeq
The contributions from the charged scalars to the magnitude of $h\rightarrow \gamma\gamma$ are proportional to the square of their electric charge, with the contributions from $H_p^{\pm\pm\pm}$ and $H_m^{\pm\pm\pm}$ being nine times
larger those from $H_{p}^{\pm}$ and $H_m^{\pm}$. Hence, the $\Gamma({h\rightarrow \gamma\gamma})$ in this model is dominated
by the contributions from $H_p^{\pm\pm\pm}$ and $H_m^{\pm\pm\pm}$.

For the case of $\lambda_4>0$, the interference of contributions from $H_{p,m}^{Q}$ with that from $W$ loop is destructive as dictated in equation \ref{ffun}. As the increase of $\lambda_4$, the contributions from charged particle $H_p^{\pm\pm\pm}$ and $H_m^{\pm\pm\pm}$ become large even dominant. For $\lambda_4>0.12 $ (assuming $70\ \rm{GeV} <m_{H_{p,m}^{Q}}< 90\ \rm{GeV}$), $\mu_{\gamma\gamma}$ could be larger than $1.5$.
Compared to the case in Higgs triplet model where a value of $\mu_{\gamma\gamma}>1$ is difficult to obtain unless in a narrow region where the trilinear coupling $\lambda>5$ and $m_{H_p^{\pm\pm}}<200$ GeV \cite{hrrHTM}, a large enhancement $1.5<\mu_{\gamma\gamma}<2$ is easier to get in small $\lambda_4$ region in S7M. This is because more charged scalars
, especially, the triply charged scalar, are introduced here which contribute significantly to the width.

In contrast, the interference of contributions from new charged particles and with the contribution from $W$ loop
is constructive for $\lambda_4<0$. As shown in the Fig.~\ref{Fig:hrr}, a moderate enhancement can be achieved in a large parameter space, e.g. enhancement factor $\mu=1.5$ for $\lambda_4=-0.05, -0.5, -5$ with $m_{H_{p,m}^{Q}}=108\ \rm{GeV},\ 311\ \rm{GeV},\ 975\ \rm{GeV}$, respectively. However, the current sensitivity of $h\rightarrow \gamma\gamma$ in the LHC searches constrain significantly on this $\lambda_1<0$ region. For a point with very small value of $m_{H_{p,m}^{Q}}$ and very large $\lambda_4$, the enhancement is too large to be consistent with LHC search results.

\begin{figure}[htb]
\begin{center}
 \includegraphics[scale=0.65]{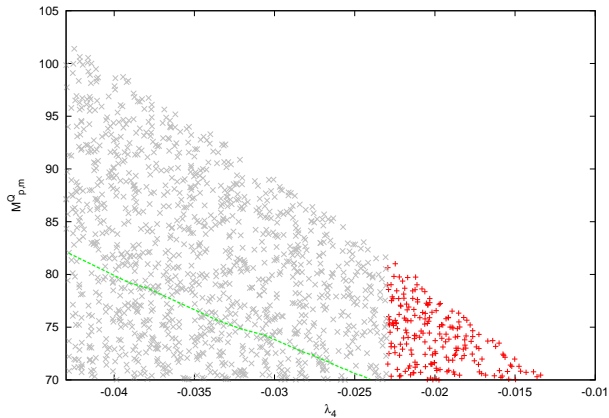}
\end{center}
\caption{The scatter plot of $\lambda_4$ versus charged scalar mass with enhancement factor $\mu_{\gamma\gamma}$ in the range $1.5\sim 3$. The points above the green dashed line have $\mu_{\gamma\gamma}$ in the range $1.5\sim 2$ while the points below the line have $\mu_{\gamma\gamma}$ in the range $2\sim 3$.}
\label{Fig:DMandRatio}
\end{figure}

Furthermore, we are more interested in the small $\lambda_4$ region since the dark matter relic abundance and direct detection favor this region, as shown in Fig.~\ref{Fig:DMandRatio}. For grey points in the region of $-0.043 <\lambda_4< -0.023$ and $70~{\rm GeV}< m_{H_{p,m}^Q}< 102~{\rm GeV}$, we get an enhancement factor of $1.5\sim3$ in $h\rightarrow \gamma\gamma$ channel. For these points, the dark matter mass is in the range $[64~{\rm GeV}, ~67~{\rm GeV}]$ consistent with the $2\sigma$ constraint of XENON 100. The red points are below the central value of the XENON 100 constraint.

\section{Collider Signatures}
In this section, we study the collider phenomenology of this model.
The scalar sector includes the SM like Higgs $h$, the CP odd neutral Higgs $A$ and the CP even neutral Higgs $H^0$ in neutral sector, and charged scalars $H_{p,m}^{\pm}$, $H_{p,m}^{\pm\pm}$, $H_{p,m}^{\pm\pm\pm}$.
The lightest neutral component of septuplet $H^0$ introduced in the S7M is a stable particle and thus the
dark matter candidate, whose signal at colliders is missing energy. Because there is no mixing among the $h$, $H^0$ and $A$, the Higgs boson $h$ couplings to SM particles are not changed except the effective couplings of
$h\gamma\gamma$ and $hZ \gamma$.

The singly charged Higgs and the doubly charged Higgs are the featured particles in the two Higgs doublet model (2HDM) and the triplet Higgs model (THM) respectively.
There are associated extensive studies \cite{2HDM,HTM}.
Current searches for singly charged Higgs mainly focus on processes $t\bar{t} \rightarrow H^{\pm}W^{\mp}b\bar{b}$ and
$t\bar{t} \rightarrow H^{\pm}H^{\mp}b\bar{b}$ \cite{LEP,ATLASH+,CMSH+} \footnote{Ref.~\cite{Yang} found jet substructure can improve search potential for heavy charged Higgs at the LHC.}. For the doubly charged Higgs $H^{++}$, the general search considers its decays to same-sign leptons with same flavor or different flavor \cite{ATLASH++,CMSH++}.

However, these experimental constraints do not apply to the scalars in the S7M. Unlike the case in the general 2HDM and THM where Yukawa couplings are introduced and the charged scalars ($H^{++}$ or $H^{+}$) can decay into a fermion pair, the Yukawa couplings in this model are naturally forbidden naturally. Furthermore, in the general 2HDM and THM, the neutral component $H^0$ of the additional Higgs doublet or the Higgs triplet gets vacuum expectation value (VEV). As a result,lt in the charged Higgs ($H^{++}$ or $H^+$) can decay to a gauge boson pair ($W^+W^+$ or $W^+Z$ ). In contrast, the neutral component of the septuplet does not get a VEV in this model.

The direct constraints in this model mainly similar to supersymmetry searches, since the collider signatures for $A$, $H^0$, $H_p^{\pm}$, $H_m^{\pm}$ are similar to those of neutralinos and charginos in the Minimal Supersymmetric Standard Model (MSSM). $H^0$ in this model appears as missing energy at colliders, similar to $\chi_1^0$ as the lightest supersymmetric particle(LSP) in the MSSM. The CP-odd neutral particle $A$ and the charged Higgs $H_p^{\pm}$ are similar to the $\chi_2^0$ and $\chi_1^{\pm}$ respectively.
The searches for neutralino and chargino at the LHC \cite{ATLASX+X0} via $pp\rightarrow \chi_1^0\chi_2^0$ and $pp\rightarrow \chi_1^0\chi_1^+$, $\chi_1^+\chi_1^-$ can be used to set limits on $m_{H^{\pm}}$, $m_A$ and $m_{H^0}$ in the S7M. Generally, the experimental analyses are focused on decay channels $\chi_2^0\rightarrow \chi_1^0 Z^{*}\rightarrow ff\eslash_T$ and $\chi_1^{\pm} \rightarrow \chi_1^0 W^{\pm *} \rightarrow ff \eslash$. Especially  $l^+l^- \eslash_T$ and $l^+l^-l^+ \eslash_T$ are "Golden" channels for the SUSY search.


There are some differences between the S7M and the MSSM. The existence of $H_{p,m}^{\pm\pm\pm}$ is a distinctive feature of the S7M. These triply charged scalars can only appears in a high dimension representation of $SU(2)_L$ and the detection of these particles is the smoking gun of the S7M. The main decay channel for triply charged Higgs is $H_{p,m}^{\pm\pm\pm}\rightarrow H_{p,m}^{\pm\pm} W^{\pm*} \rightarrow H_{p,m}^{\pm\pm}ff'$ in this model.
However, the mass splitting among the charged particles are not very large. The suppression of the phase space and weak
couplings results in a displaced vertex in the detector.
At the LHC, $H_{p,m}^{+++}$ can be produced via processes $pp\rightarrow \gamma^*,Z^*\to H_{p,m}^{+++} H_{p,m}^{---}$ and $pp\rightarrow W^*\to H_{p,m}^{+++} H_{p,m}^{--}$. For a $H_{p,m}^{+++}$ with mass around 100 GeV, the cross section is estimated to several fb. A detailed analysis on collider phenomenology of this model is left for future study.

\section{Conclusions}
In this paper, we studied the extended SM Higgs with two septuplets. Without loss of generality, we assume one septuplet, $\Phi_7$, is much heavier. The new particles include the additional CP-even scalar $H^0$, the CP-odd scalar $A$ and the charged scalars $H_{p,m}^{\pm}$, $H_{p,m}^{\pm\pm}$, $H_{p,m}^{\pm\pm\pm}$. The lighter one of the neutral components $H^0$ could be a good dark matter candidate. We explored the parameter space, taking into account the relic density constraint from WMAP and direct detection constraint. Furthermore, we study the contributions from new charged particles to Higgs decay in di-photo channel $h\rightarrow \gamma\gamma$ to address the enhancement of the decay rate.
It is novel that this model provides a good dark matter candidate of mass around 70 GeV and accommodates the enhancement factor around $1.5\sim 2$ in $h\rightarrow \gamma\gamma$ channel. The collider signature of this model has also been discussed, while the detailed analysis will be left for the future study. It is expected that this model could be tested with LHC and dark matter detection in the near future.

\vspace{4mm} \textbf{Acknowledgments}\\

We would like to thank KITPC for great hospitality during this work.
We would also like to thank Qi-Shu Yan, Gui-Yu Huang and Tao Liu for helpful discussions.
This work was supported in part by the National Natural Science Foundation of China under Grants No.11175251,
and in part by Postdoctoral Science Foundation of China under Grants 2012M510248 and SJTU Postdoc Fellowship.

\end{document}